\documentstyle[12pt,epsfig]{article}
\newcommand{\blankline}{\vskip .3cm}
\newcommand{\f}{\begin{equation}}
\newcommand{\ff}{\end{equation}}
\newcommand{\be}{\begin{equation}}
\newcommand{\ee}{\end{equation}}
\newcommand{\bea}{\begin{eqnarray}}
\newcommand{\eea}{\end{eqnarray}}
\begin{document}
\centerline{\LARGE Generalized Lorentz invariance with an invariant energy
scale}
\blankline
\blankline
\rm
\centerline{Jo\~ao Magueijo${}^\#$ and Lee Smolin${}^{* }$}
\blankline
\centerline{\it ${}^\#$The Blackett Laboratory,Imperial
College of
Science, Technology and Medicine }
\centerline{\it South Kensington, London SW7 2BZ, UK}
\centerline{\it ${}^*$Perimeter Institute for Theoretical Physics}
\centerline{\it Waterloo, Canada  N2J 2W9}
\centerline{\it  \ \ and }
\centerline{\it Department of Phyiscs, University of Waterloo}

\blankline
\blankline
\blankline
\blankline
\centerline{July 22, 2002}
\blankline
\blankline
\blankline
\blankline

\centerline{ABSTRACT} The hypothesis that the Lorentz
transformations may be modified at Planck scale energies is
further explored.  We present a general formalism for theories which
preserve the
relativity of inertial frames with a non-linear action of the Lorentz
transformations on momentum space.
Several examples are discussed in which the speed of light varies with
energy and elementary particles have a maximum momenta and/or
energy.  Energy and momentum conservation are suitably generalized
and a proposal is made for how the new transformation laws apply
to composite systems. We then use these results to explain the ultra
high energy cosmic ray anomaly and we find a form of the theory that
explains the anomaly, and leads also to a maximum momentum and a
speed of light that diverges with energy.
We finally propose that the spatial
coordinates be identified as the generators of translation in
Minkowski spacetime. In some examples this leads to a commutative
geometry, but with an energy dependent Planck constant. \vfill
\blankline

\blankline
lsmolin@perimeterinstitute.ca,\ \  j.magueijo@ic.ac.uk
\eject
\tableofcontents
\eject
\vfill

\section{Introduction}

Several experimental and theoretical developments point to the
possibility that the usual relation between energy and momentum that
characterize the special theory of relativity,
\f
E^2 = p^2 + m^2
\label{usual}
\ff
may be modified at Planck scales.   Among these is the observed
 threshold
anomalies in ultra high-energy cosmic ray
protons\cite{review,crexp}, and possibly also Tev
photons\cite{gammaexp}.
    As pointed out first by Amelino-Camelia and Piran \cite{piran},
these can be
    explained by modifications of the energy momentum relations of the
    form,
    \f\label{moddisp}
    E^2 = p^2 + m^2 + \lambda E^3 + ...
    \label{modified}
    \ff
    where $\lambda$ is of the order of the Planck length.

Such a modified energy-momentum relationship
leads to further predictions which are
falsifiable with planned experiments. Among these is an energy
dependent speed of light, observable (with $\lambda$ of the order
of the Planck length) in planned gamma ray observations
\cite{amelinovsl}. An energy dependent speed of light may also
imply that the
    speed of light was faster in the very early universe, when the
    average energy was on comparable to Planck energies~\cite{ncvsl}.
As pointed
    out by Moffat~\cite{moffat93},
and Albrecht and Magueijo~\cite{am}, such an effect could
    provide an alternative solution to the horizon problem and other
    problems addressed by inflation.  Such modified dispersion
    relations also may lead to corrections to the predictions of
    inflationary cosmology, observable in future high precision
measurements of the CMB spectrum\cite{pogo}. Finally a modified
dispersion relation may lead to an explanation of the dark energy,
in terms of energy trapped very high momentum and low-energy
quanta, as pointed out by Mersini and collaborators\cite{laura}.

These effects open up the possibility of testing hypotheses about
Planck scale physics by more than one window, in the present and
near future.  Indeed, the fact that the hypothesis that the
energy-momentum
relations receive corrections of the form of (\ref{moddisp})
is experimentally testable, with $\lambda$ on the form of the Planck
scale, is alone sufficient reason to consider it.

However there are also theoretical issues which
motivate such a modification.  Several calculations\cite{loopcalc}  in
loop
quantum gravity in fact predict modified
    dispersion relations of the form (\ref{modified}).  That they do so
    is not surprising for, from the point of view of the quantum theory
    of gravity, global Lorentz invariance is no more than an accidental
    symmetry of the ground state of the classical limit of
    the theory\footnote{Neglecting the cosmological constant.}.
    Thus, it is to be expected that corrections to consequences of
    Lorentz invariance appear as quantum gravitational effects, which
    is to say as corrections to the laws of special relativity
    which are suppressed by $l_{Planck}$.

At the same time, there is a simple reason to be skeptical that the energy
momentum relations may receive modifications of the form of
(\ref{moddisp}). Such a modification contradicts the transformation
laws of special relativity, according to which energy and momentum
transform according to the Lorentz transformations, so as to preserve
the Minkowski metric. Lorentz invariance is generally assumed to be a
consequence of the oldest and most reliable principle in all
of dynamics, which is the relativity of inertial frame.  One may then
worry that were  the hypothesis
(\ref{moddisp}) confirmed experimentally, that event would signal
that after all there is a preferred
frame of reference in nature, in contradiction to the last
400 years of progress in science.

Very recently several people have realized that this worry is
unnecessary\cite{doub,leejoao,nv}.
It is possible to keep the principle of the  relativity of
inertial frames, and simply modify the laws by which energy and
momenta measured by different inertial observers are related to each
other.  By adding non-linear terms to the action of the Lorentz
transformations on momentum space, one can maintain the relativity
of inertial frames. The quadratic invariant is replaced by a non-linear
invariant, which in turn leads to a modified dispersion
relations of the form of (\ref{moddisp}).

There is indeed a simple argument that suggests that such a
modification may be necessary.  In quantum theories of gravity such as
loop quantum gravity, the Planck length plays the role of a
threshold below which the classical
picture of smooth spacetime geometry gives way to a discrete quantum
geometry. This suggests that the Planck length plays a role analagous
to the atomic spacing in condensed matter physics.  Below that
length there is no concept of a smooth metric. It is then not
surprising if quantities involving the metric, such as the quadratic
invariant, receive corrections of order of the Planck length.

However this raises
a problem.  Lengths are not invariant under Lorentz
transformations, so one observer's threshold will be perceived to
be set in at a different length scale than
another's\footnote{For another view of this
apparent paradox, however, see \cite{rov}.}.

Alternatively, various
hypotheses concerning quantum gravity \cite{rovelli,carlip} and
string theory \cite{pol,stringref}
suggest that the geometry of spacetime is in fact
    non-commutative. In such a modification the spacetime
    coordinates may no longer commute and there may be modified
    energy-momentum relations. This in turn suggests a
    deformation of the Poincare symmetry of flat spacetime, one
    example of which is given by the  $\kappa$-Poincare symmetry
    discussed, for example, in \cite{kappa1,majid,kappa3}.
In all these proposals, the non-commutativity is measured by a
parameter which has the dimensions of a length. Again, we can ask how
it is that all observers agree on the scale at which
non-commutativity appears, given that lengths normally are not invariant
under Lorentz transformations.

These paradoxes may be resolved if the lorentz transformations may be
modified so as to preserve a single energy or momentum scale. Then all
observers will agree that there is an invariant energy or momentum
above which the picture of spacetime as a smooth manifold breaks down.
Because there are then two constants which are preserved, this
proposal has been called ``doubly special relativity\cite{doub,nv}.''

To summarize, the various experimental and theoretical issues we
have mentioned lead us to ask whether it is possible to modify
the principles of physics so that all of the following requirements
are met.

\begin {enumerate}
\item The relativity of inertial frames, as proposed by
Galileo, Descartes, Newton and Einstein, is preserved.

\item There is nevertheless an invariant energy scale
$E_P=\lambda^{-1}$, which is of the order of the Planck scale.

\item The threshold for UHECRs should be
increased as suggested by experiment\footnote{The situation for
the Tev photon threshold is not as convincing, so we do not yet
impose it as a requirement.}.

\item The theory should exhibit a varying, preferably diverging, speed
of light, at high energies.

\item The theory should have a maximal momentum, corresponding to
the granular nature of space.
\end{enumerate}

The main question this paper raises is whether it is possible to modify
the
principles of special relativity so that all five of these requirements
are met.
The main result of this paper is that the answer is affirmative.

To explain the viewpoint we take, we can start by emphasizing that
    no matter how quantum mechanical, non-commutative or deformed the
    geometry of
    spacetime may become, if the principle of the relativity of
    inertial frames is to hold, the transformations between
    measurements made by inertial observers must satisfy the group
property.
    Furthermore, the group must be a six parameter extension of the
    rotation group, with the three additional parameters going over
    into the boosts of special relativity whenever quantities of the
    order of the Planck scale can be ignored. However,
    as we argued in \cite{leejoao} the only group with these
    properties is the Lorentz group itself. Hence, the only possibility
    to achieve all of these conditions is through a non-linear action
    of the ordinary lorentz group on the states of the theory.

    In a recent paper, we proposed this viewpoint, and we proposed a
    class of theories in which the lorentz transformations act
    non-linearly on momentum space. We studied in some detail a
    a simple example\cite{leejoao} of a theory in which the
    action of the lorentz group on momentum space was made non-linear,
    in such a way that the Planck energy became an invariant.
    However, as we will demonstrate below, this example does not satisfy
all of
    the conditions just mentioned.
    Furthmore, studies by other
    authors suggested that it was not obvious how to modify the action
    of the lorentz transformations so as to achieve all these
    conditions, in particular, it appeared that theories in which the
    relativity of inertial frames is preserved may not be able to
    account for the observed threshold anomalies.

    Before going further, we want to emphasize that the proposal
    to modify the action of the lorentz transformations on momentum
    space was not original with us.
    Modifications of special relativity in which the action of the
    transformations are non-linear have been considered by a number
    of authors\cite{fock,nv,amelinovsl,doub,kappa3,kappa1,majid}.
    To our knowledge
    the earliest such proposal
    is by Fock\cite{fock} and related proposals have been considered
    earlier also in \cite{kappa3,majid,kappa1,nv,amelinovsl,doub}.
    We consider that our contribution is
    mainly to take a phenomenological point of view in which we insist
    that the modifications of special relativity are to be treated in
    the most general way possible, so as to allow nature to teach us
    if and how the relativity of inertial frames is realized in a
    fundamental theory.  While it may in the future happen that a
    fundamental quantum theory of gravity makes predictions for
    the exact form of a modified dispersion relation or action of
    the lorentz transformations, we want to avoid too hastily following
    any particular mathematical hypothesis about the structure of
spacetime
    to the exclusion of others.

    Recently a number of authors have contributed to the study of
    such theories, discussing many aspects which we are not able
    to consider here\cite{contr,granik,spin}. However we are able
    to address a number of issues which concern the
whole class of theories in which the lorentz transformations act
non-linearly on momentum space.

\begin{itemize}

    \item{}It is not hard to see that if one adds momenta and energy
linearly,
    as we normally do in physics,
    the conservation of momentum is inconsistent with the new
    nonlinear action of the lorentz group on momentum space. Is it
    then possible that energy and momentum conservation is
    maintained, but with these laws also becoming non-linear?

    \item{}There is no experimental reason why the energy and momenta
    of elementary particles
    cannot be bounded by  the Planck energy, but this is certainly not the
case
    for macroscopic systems.  Thus, the transformation laws must
    distinguish elementary from composite systems in such a way that
    the macroscopic bodies can have Planck energies and momenta while
    transforming and moving according to the usual laws of special
    relativity.

    \item{}If a modified energy momentum relation is the explanation
    for the observed threshold anomalies, there appears to be a
    necessity that these modifications are significant already at the
    scales of the thresholds themselves, which are on scales of
    $Tev$'s for the photons and $10^{11} Gev$ for the protons, i.e.
    small compared to
    the Planck energy. At the same time, there is no observed energy
    dependence of the speed of light, seen in gamma ray observations,
    to scales up to $10^{-3} E_{Planck}$, which is in fact much
    higher than the scale of the problematic thresholds. So how could
    a theory resolve the problem of the threshold anomalies, without
    at the same time causing an energy dependent speed of light at
    scales which are already ruled out by experiment?

    \item{}Before the present viewpoint was formulated a number of
    authors arrived at modified lorentz transformations by
    investigating the hypothesis that geometry becomes
    non-commutative, so that spacetime coordinates no longer commute.
    A beautiful example of such a theory is the $\kappa$-Poincare
    symmetry of $\kappa$-deformed Minkowski spacetime, which is
    indeed a non-commutative geometry\cite{majid,kappa3}.
    However, it is important to ask
    whether the non-commutativity of spacetime coordinates is a
    necessary consequence of modifying the action of the lorentz
    transforamtions on momentum space, so as to have an invariant
    Planck energy and/or momentum, or whether one can achieve such a
    modification in the context of a commutative spacetime geometry.

\end{itemize}

In this paper we will resolve all of these issues,
in the course of showing that all $5$ of our conditions can be met.

The plan of this paper is as follows. In Section~\ref{gen} we
start by noting that the procedure defined in \cite{leejoao} is
not unique: there are other possible non-linear realizations of
the Lorentz group in momentum space, associated with different
operators in place of $U(p_0)$ (defined in \cite{leejoao}). Each
of these leads to different modified invariants and hence to
different dispersion relations for massive and massless particles.
We also write down the necessary and sufficient
conditions for a general non-linear action to display an invariant
energy scale, and a maximal momentum.  We  place conditions upon
$U$ so that the group property of the action is preserved and
highlight the emergence of a preferred frame should these
conditions be violated.

In section 3 we discuss several examples, and show how the variable speed
of
light theories discussed in cosmology\cite{moffat93,am,ncvsl} and
the $\kappa$-Poincare group fit into the general framework
introduced here.

Then, in Section 4 we explain how energy and momentum
conservation are maintained in these theories, a matter closely
related to the definition of momenta addition.  This allows us
in section 5 to examine
the kinematics of UHECRs and gamma rays, placing constraints upon
the possible realizations of our theory which can explain current
observations. We find that standard deformations can only explain
the threshold anomalies with a {\it negative} Planck energy, of
the order of $10^{11}$ Gev. However we exhibit one class of
dispersion relations (and an associated non-linear Lorentz action)
where this problem does not exist.

Finally in Section~\ref{real}, we point out that because the
Lorentz group acts non-linearly on momentum space, the action on
space-time coordinates is also non-trivial. We propose that in
quantum theory the space and time coordinates are to be
    defined to be generators of translations in momentum space. In
    contrast to other definitions, this leads to a commutative
    spacetime geometry. But the commutation relations between
    position and momentum become energy dependent, leading to a new
   energy dependent modification of the uncertainty relations.

 \section{Generalized non-linear actions and deformed
dispersion relations}\label{gen}

In our first paper\cite{leejoao}
we proposed a non-linear modification of the action
of the Lorentz group in momentum space which contains an observer
independent length  scale (denoted here by $\lambda$), and reduces
to the usual linear action at low energies. For the new proposal
the concept of metric (a quadratic invariant) collapses at high
energies, being replaced by the non-quadratic invariant
\begin{equation}\label{invariant}
 ||p||^2 \equiv {\eta^{ab}p_a p_b \over (1- \lambda p_0 )^2}
\end{equation}
The group algebra, however, is left unchanged, suggesting that the
spin connection formulation of general relativity may still be
valid (as the connection takes values in the algebra). This
remarkable feature may be traced to the fact that our modified
boost generators (and likewise for the rotation generators) may
be written in the form
\begin{equation}\label{U}
K^i = U^{-1} [p_0] L_0^{\ i} U [p_0 ]
\end{equation}
where $L_{ab} = p_a {\partial \over \partial p^b} -
 p_b {\partial \over \partial p^a}$ are the standard Lorentz generators.
For the particular boosts we have proposed we have:
\begin{equation}\label{U1}
U [p_0]\equiv \exp(\lambda p_0 D)
\end{equation}
(where $D=p_a{\partial\over \partial p_a}$ is a dilatation), or
more specifically:
\begin{equation}
U[p_0](p_a)={p_a\over 1-\lambda p_0}
\end{equation}
We note that $\lambda$ may have either sign. We expect that it
will be proportional to plus or minus the Planck length. We also
note that this is not an unitary equivalence, because, as may be
easily checked, in general $U$ is not unitary. In fact we shall be
mainly interested in singular expressions for $U$, for reasons to
be explained shortly. In addition the transformation depends on
parameter $\lambda$, and we shall at times recall this dependence
with the notation $U[p_0;\lambda]$. When there is no risk of
confusion we shall drop one or both of the variables in square
brackets. Note that the transformation induced by $U$ coincides
with the linearization procedure proposed by Judes \cite{judes}
{\it in the single particle sector.}

The choice of $U$ made in (\ref{U1}) is dictated by nothing but
simplicity and the fact that it leads to the Fock-Lorentz group
acting in momentum space\cite{fock,man,step}. Any other
non-unitary, non-linear $U$ leads to a non-trivial alternative
representation of the Lorentz group, with algebra:
\begin{equation}
[J^i, K^j]= \epsilon^{ijk} K_k; \ [K^i, K^j]= \epsilon^{ijk} J_k
\end{equation}
(with $[J^i, J^j]= \epsilon^{ijk} J_k $ trivially preserved). Any
group action generated via (\ref{U}), but with a different form
for $U$, generalizes our formalism, even if the resulting action
does not preserve $E_P=\lambda^{-1}$ (as is the case with
(\ref{U1})~\cite{leejoao}). Hence one has to face the issue of how
to decide which $U$ is the correct one. Our view is that such a
matter should be decided by experiment. To this end we note that
the most general invariant associated with the new group action is
\begin{equation}
||p||^2 \equiv \eta^{ab}U(p_a) U(p_b)
\end{equation}
from which follows a deformed dispersion relation. Dispersion
relations may then act as the experimental input into the
formalism.

Alternatively, a dispersion relation may be derived from
calculations in a theory such as loop quantum
gravity\cite{loopcalc}. If we have reason to believe that the
theory maintains the relativity of inertial frames, in spite of
the appearance of modified terms in the energy-momentum relations,
this implies that the symmetry of the ground state, corresponding
in the classical limit to Minkowski spacetime, is a non-linear
representation of the Lorentz group such as that considered here.
In this case we may read off the $U$ from the calculated
modifications to the energy-momentum relations.

Ideally, then, the formalism we are discussing can be used to
compare  experiment and theory, as well as to extrapolate between
predictions of different experimental results. Thus, we see this
formalism as being part of a phenomenology of quantum gravity effects,
as opposed to directly having fundamental significance.

\subsection{Building a general $U$-map}

Following this philosophy, we start from an hypothetical measurement
of a set of dispersion relations and from this we infer the group
action. Any isotropic dispersion relation may be written as
\begin{equation}\label{desp}
  E^2f_1^2(E;\lambda)-p^2f_2^2(E;\lambda)=m^2
\end{equation}
implying \f U \circ (E, {\bf p})=(Ef_1,{\bf p}f_2). \label{udef}
\ff $U$ defined in this way maps energy-momentum space, $\cal P$,
onto itself. In general such a map is not invertible.  For the
action of the Lorentz group to be modified according to (\ref{U})
we must have an invertible map. It must then be that there is a
sector of $\cal P$, which we will call ${\cal P}^{phys}$ such that
$U$, with range restricted to ${\cal P}^{phys}$ is invertible.
Typical examples of the restriction which
defines ${\cal P}^{phys}$ are $E < E_{Planck}$ and/or $|{\bf p }|
< E_{Planck}$.

A further condition is that the image of $U$ must include the
range $[0,\infty]$ for both energy and momentum. This is because
the ordinary Lorentz boosts $L$ span this interval, and so
$U^{-1}LU$ would not always exist otherwise. If this
condition is not satisfied the group property of the modified
Lorentz action is destroyed. If for instance $Ef_1$ does not span
$[0,\infty]$ then there is a limiting $\gamma$ factor for each
energy, a feature which not only destroys the group property but
also selects a preferred frame, thereby violating the principle of
relativity.

With these two assumptions, we can then use (\ref{U}) to construct
the modified boosts and translations on ${\cal P}^{phys}$ and
obtain the particular realization of our theory incorporating the
new results. In particular, all such theories will have an
unmodified Lorentz algebra, realized generally non-linearly on
momentum space. The restriction to ${\cal P}^{phys}$ will also
become part of the new theory.

Unless $f_1=f_2$  we obtain a theory displaying a frequency
dependent speed of light. More precisely, defining $f_3=f_2/f_1$
we have
\begin{equation}\label{speedc}
c={dE\over dp}={f_3\over 1-{Ef_3'\over f_3}}
\end{equation}
Hence our formalism may be readily adapted to varying speed of
light (VSL) theories, justifying some of the assumptions in
\cite{ncvsl}.

\subsection{A general action which preserves an energy
scale or has a maximal momentum}\label{cond}

We mentioned in the introduction that several theoretical
arguments suggest that in nature $E_{P}$ should be an invariant
under the action of the Lorentz group. We discuss here the
conditions on $U$ such that this will be the case.

Given (\ref{U}) we know that the invariants of the new
theory are the inverse images via $U$ of the invariants of
standard special relativity. But the only invariant energy in
linear relativity is the infinite energy. Hence the condition we
are looking for is
\begin{equation}\label{uep}
  U(E_P)=E_pf_1(E_P)=\infty
\end{equation}
that is, $U$ should be singular at $E_P$. In addition note that in
special relativity there are three situations in which $E=\infty$:
\begin{itemize}
\item A photon ($||p||^2=0$), for which $E=p=\infty$.
\item A particle with infinite rest mass.
\item A particle with finite rest mass moving at the speed of
light.
\end{itemize}
These are mapped by $U^{-1}$ into 3 distinct types of objects that can
have
the (invariant) Planck energy: those with zero, infinite, and
finite mass, respectively photons, particles and something we may
call infinitons. The latter have the
property that, like photons, they cannot be boosted to a rest
frame; however they are not zero mass objects. The second have the
property that their momentum can only have two values: zero or the
Planck momentum.  These objects will generally mark the boundaries
of ${\cal P}^{phys}$. As in the case
of the limiting velocity of the speed of light in ordinary
special relativity, whether they are limiting idealizations,
or real physical cases, depends on the dynamics of the
particular theory.

The condition for the existence of a maximal momentum is simply
that $Ef_1/f_2=E/f_3$ has a maximum. Hence we note that the
conditions for a varying speed of light and for the existence of a
maximum momentum are related, and indeed one may show that
existence of a maximum momentum implies that the speed of light
must diverge at some energy.

\section{Some examples}

We now turn to discuss several examples of theories that
meet various of the requirements we posited.
In Section~\ref{ex1} and \ref{ex11} we discuss
examples of VSL theories. Another interesting exercise (performed
in Section~\ref{ex2}) consists of using the dispersion relations
associated with the $\kappa$-Poincare group to build a realization
of our theory.

\subsection{A VSL dispersion relation}\label{ex1}
The varying speed of light scenario \cite{moffat93,am} is an
interesting alternative to cosmological inflation. It was found in
\cite{ncvsl,ncinfl} that some deformed dispersion relations (such as
the ones in \cite{amel,amel1,majid}) might lead to a realization
of VSL (and even inflation).

The dispersion relations for massless particles were written in
\cite{ncvsl,ncinfl} in the form $E^2-p^2f^2(E)=0$, failing to
define fully $f_1(E)$ and $f_2(E)$. However as an example let us
consider the case of \cite{amel1} with $f_2=f=1+\lambda E$, and
$f_1=1$. This model is known to have an energy dependent speed of
light $c(E)={dE\over dp}=(1+\lambda E)^2$; also all momenta must
be smaller than the maximum momentum $p=\lambda^{-1}$, which can
only be reached by photons with infinite energy. We note that if
the theory is to provide a solution to the horizon problem,
independent of inflation, we require $\lambda >0$. Even though the
image of $U$ does not span $[0,\infty]$ if we restrict ourselves
to positive energies, it does so in $E\in[-\infty,\infty]$, so the
group property is preserved for this proposal.

Following the procedures described above we arrive the following
transformation laws for photons:
\begin{eqnarray}
E'&=&\gamma (1-v)E\\
p'&=&{\gamma(1-v)p\over 1+\lambda p(\gamma(1-v)-1)}
\end{eqnarray}
Although no energy remains invariant, the Planck momentum
$p=\lambda^{-1}$ is an invariant and is also the maximal momentum.
The gravitational redshift formula is unmodified in this theory,
but expressions for phenomena involving exchange of momentum will
be different.

For massive particles we find that we still have that
$E_0=m_0c^2$, that the mass still transforms like $m=m_0\gamma$
and that in any frame $E=mc^2$. However the general expression for
the momentum is now \be p={mv\over 1+\lambda m} \ee showing that
for massive particles we must have $p<p_{max}=\lambda^{-1}$.

Similar expressions may be derived for other VSL models considered
in the literature.

\subsection{A second VSL theory}\label{ex11}

A second VSL theory is obtained  by choosing
\f U= e^{-\lambda E^2
\partial /\partial E}
\ff leading to the energy momentum relation \f \label{vsl2}{E^2
\over (1+\lambda E)^2} - p^2 = m^2 \ff This results in the same
modification of the speed of light for photons as the first
example, but differs in the energy-momentum relation for massive
particles. Note that in this case there is an invariant energy
scale, which with the notation used is $E=-\lambda^{-1}$, that is,
it's negative for $\lambda>0$.

A more general set of isotropic energy momentum relations
may be derived from the choice,
\f
U= e^{ g_1(E) E \partial /\partial E + g_2(E) p_i \partial /
\partial p_i}
\ff We thus see the need for experiment, or further theoretical
considerations, to fix the high energy behavior of the action of
the lorentz transformations on momentum space.

\subsection{The $\kappa$-Poincar\'e group}\label{ex2}

The $\kappa$-Poincar\'e group is a quantum deformation of the
usual Poincar\'e group \cite{kappa1,jur,kappa3}, which we now show can be
reinterpret in our formalism.  It leads to dispersion relations of
the form (\ref{desp}) with \bea\label{kappa}
f_1&=&{\sinh(\lambda E)\over \lambda E}\\
f_2&=&\exp(\lambda E) \eea from which a $U$ can be read off
according to our prescription. It leads to modified boost
generators: \be F_i={e^{\lambda E}\over \cosh\lambda
E}[p_i\partial_E-\lambda p_iD] +{\sinh\lambda E\over \lambda
e^{\lambda E}}\partial_{p_i} \ee and it can be checked that these
satisfy the standard Lorentz algebra.

Exponentiation reveals the finite Lorentz transformations: \bea
E'&=&\lambda^{-1} \sinh^{-1}(F)\\
p'_z&=&{p_ze^{\lambda E}-v\lambda^{-1}\sinh(\lambda E) \over
F +{\sqrt{F^2 +1}}}\\
p'_x&=&{p_x\over
F +{\sqrt{F^2 +1}}}\\
p'_y&=&{p_y\over
F +{\sqrt{F^2 +1}}}\\
F(E,p_z)&=&\gamma(\sinh(\lambda E)-vp_z\lambda e^{\lambda E}) \eea
For photons these reduce to the Doppler shift formula: \be
E'=\lambda^{-1} \sinh^{-1}(\gamma(1-v)\sinh\lambda E) \ee For
massive particles we have the relation \be E=\lambda^{-1}\log
[(\lambda m)+{\sqrt{(\lambda m)^2+1}}] \ee with $m=\gamma m_0$.

Note that although the theory we have written down has the same
dispersion relations as those of the $\kappa$-Poincare group, this may
not necessarily imply that the structure of spacetime must be
assumed to be non-commutative.  Our
theory is not based on a quantum
deformation of the Poincare group, but merely a non-linear
realization of the undeformed Lorentz group.  This is true even in
the case in which the dispersion relations are the same as derived
from the $\kappa$-Poincare group. We will see below how the
space-time coordinates may be introduced, in a way that does not
require the introduction of non-commutative spacetime geometry.

\section{Composite systems and conservation laws}\label{multip}

Once we accept the possibility of non-linear transformation laws,
we soon discover that kinematic relations valid for single
particles need not be true for composite systems (this is
certainly the case with the transformation laws themselves). In
fact we are left with an ambiguity concerning how
momenta are added and how  composite quantities transform. To some extent
this is a desirable feature: non-linearity appears to build into
the theory the concept of elementary particle, clearly
differentiating between them and composites. In any case, as we
mentioned in the introduction, such a distinction is necessary for
theories in which energy or momentum of elementary particles are
bounded.

\subsection{Composite systems}

One has to tread gingerly when defining the multi-particle
sector, as
theories predicting deformed dispersion relations often have
ill-behaved multi-particle sectors\footnote{We thank G. Amelino-Camelia
for bringing this point to our attention.}. For instance, a
straightforward addition rule is the following:
\begin{equation}\label{add1}
p_a^{(12)}=U^{-1}(U(p^{(1)}_a)+U(p^{(2)}_a))
\end{equation}
and likewise for larger collections of particles, $p^{(1...n)}$.
This is the simplest composition map, and it does lead to
the conservation of energy and momentum, as may be easily checked.

However there is a problem with this law. As may be checked,
with this definition the composite momenta transform according to
the same non-linear equations as the momenta of the constituents.
This quickly leads to inconsistencies, for instance, it implies
that composite momenta  satisfy the same deformed dispersion
relations as single particles (with $m_{(1...n)}=m_1+...+m_n$). For
the choice (\ref{U1}) this implies that a set of particles
satisfying $E\ll E_P$ can never have a collective energy larger
than $E_p$. This is blatantly in contradiction with observation:
macroscopic collections of
objects with $E\ll E_P$ quite often have energies far in excess of
$E_P$, and in fact satisfy to good approximation undeformed
dispersion relations. A severe inconsistency has arisen, traceable
to definition (\ref{add1}).

There are several ways around this problem. One was proposed in
\cite{judes}, here we describe another.
We have noted before that the
transformation $U$ depends on $p_0$ but also on parameter
$\lambda$, and we restore the latter dependence with notation
$U[p_0;\lambda]$. The idea is now that a system of $n$ elementary
particles should satisfy kinematical relations obtained from
a map $U[p_0;\lambda/n]$, that is,
a map for which the Planck energy $E_P=\lambda^{-1}$
is replaced by $nE_P$. We can therefore define:
\begin{equation}
p^{(12)}=U^{-1}\left[p_0;\lambda/2
\right]((U[p_0;\lambda ](p^{(1)})+U[p_0;
\lambda](p^{(2)}))
\equiv p^{(1)}\oplus p^{(2)}
\end{equation}
This defines a new, generally nonlinear, composition law for
energy and momenta, which we denote by $\oplus$ to indicate that
it is not ordinary addition. In general
\begin{equation}\label{add2}
p^{(1...n)}=U^{-1}[p_0;\lambda /n ](U[p_0; \lambda
](p_1)+...+U[p_0;\lambda ](p_n))
\end{equation}
With this definition a system of $n$ particles satisfies a system
of transformations obtained from $U[p_0;\lambda/n]$ via (\ref{U}),
equivalent to the usual ones but replacing $\lambda $ with $\lambda /n$.
As a result, the collective momentum $P^{(N)}=p^{(1...n)}$ satisfies
deformed dispersion relations with $\lambda $ replaced by $\lambda /n$.
This can never lead to inconsistencies, because if all $n$ particles
of a system have sub-Planckian energies then the total will still be
sub-Planckian, in the sense that $E_{tot}\ll nE_P$. We have threfore
circumvented the paradox described above.

For instance, in the case where $U$ is given by (\ref{U1}) the
addition rule for $N$ particles becomes:
\begin{equation}\label{addU1}
{p^{(N)}_a\over 1- {\lambda p^{(N)}_0\over N}}=\sum_i {p^{(i)}_a\over
1-\lambda p^{(i)}_0}.
\end{equation}
Interestingly, for sets of particles with identical energies and
momenta, this reduces to plain additivity.

The composite momentum satisfies the dispersion relations:
\begin{equation}\label{invariant1}
 ||p^{(N)}||^2 \equiv {\eta^{ab}p^{(N)}_a p^{(N)}_b
 \over {\left(1- {\lambda p^{(N)}_0\over N } \right)}^2}=M^2
\end{equation}
and the transformation laws are now:
\bea \label{fltransp}
p^{(N)'}_0&=&{\gamma\left(p^{(N)}_0- vp^{(N)}_z \right)
\over 1+{\lambda \over N} (\gamma -1) p^{(N)}_0  -
{\lambda\over N}\gamma v p^{(N)}_z }\\
p^{(N)'}_z&=&{\gamma\left(p^{(N)}_z- v p^{(N)}_0\right)
\over 1+{\lambda\over N} (\gamma -1) p^{(N)}_0  -
{\lambda\over N}\gamma v p^{(N)}_z }\\
p^{(N)'}_x&=&{p^{(N)}_x
\over 1+{\lambda\over N} (\gamma -1) p^{(N)}_0  -
{\lambda\over N}\gamma v p^{(N)}_z }\\
p^{(N)'}_y&=&{p_y^{(N)}
\over 1+{\lambda\over N} (\gamma -1) p^{(N)}_0  -
{\lambda\over N}\gamma v p^{(N)}_z }
\eea
If $p_0^{(i)}\ll E_P=\lambda^{-1}$ for all $i$, we find that
(\ref{addU1}) reduces to standard addition and $p^{(N)}$
transforms according to the usual Lorentz transformations
and satisfies quadratic dispersion relations.

More generally the proposal (\ref{add2}) for momenta addition has
the following $U$-independent properties. It's commutative but
non-associative, as expected from any addition law incorporating
the concept of elementary particle. If each elementary particle
satisfies $E\ll \lambda^{-1}$, then energy and momentum are
approximately additive and the composite momenta approximately
satisfy all relations and laws of linear special relativity. Hence
our defintion avoids the pathologies of the choice (\ref{add1}).
However if a single particle within a collection is Planckian (so
that its invariant is $m=\infty$), then the full collection (say,
of $N$ particles) is Planckian (it has energy $E_N=NE_P$.) This
can be proved by noting that the collection also has infinite
total mass. This feature is physically acceptable.

The addition law (\ref{add2}) can be generalized to:
\begin{equation}\label{add3}
p^{(1...n)}=U^{-1}[p_0;f(n,\lambda )](U[p_0;\lambda]
(p^{(1)}+...+U[p_0;\lambda ](p^{(n)})
\end{equation}
where $f(n,\lambda)$ can be any function leading to suitable limiting
properties. Then (\ref{add2}) is a particular case of (\ref{add3})
with $f(n,\lambda)=\lambda/n$.

This proposal solves the problem of how macroscopic bodies
transform at a cost, which is that for an observer
to transform the energy and momenta of a system from their
measurements to those made by an observer moving with respect to them,
they must know if the system is elementary or composite and, if
composite, how many quanta it contains.  The idea that kinematics
should make a distinction between elementary and composite systems
is new to physics, but we would like to suggest that this is not
necessarily a reason to abandon it. Instead, it is possible that this
is a feature of a classical or quantum mechanics of fundamental
particles.  For example, a theory that makes such a distinction may
be able to resolve the measurement problem because it has an objective
way to distinguish macroscopic bodies from fundamental particles.

\subsection{Energy and momentum conservation}

Our theory conserves energy and momentum because it is space-time
translation invariant. However the theory is non-linear, and the
$p_a$ of a system of two particles is not the sum $p_a^1+p_a^2$, a
matter which filters into the definition of energy-momentum conservation.
This point was made in \cite{judes}, and is closely related the
definition of momenta in the multiparticle sector discussed in the
previous section. We find that in the same way that the transition
from Galilean to special
relativity destroys the additivity of speeds, the transition from
linear to non-linear relativity destroys additivity of energy and
momentum themselves. Hence energy-momentum conservation, say for a
2-body collision, can now be written as
\begin{equation}
p_a\oplus q_a=p'_a\oplus q'_a
\end{equation}
where unprimed/primed variables refer to energy-momenta
before/after the collision. For instance for the choice (\ref{U1})
we have:
\begin{equation}
{p_a\over 1-{\lambda p_0}}+{q_a\over 1-{\lambda q_0}}={p'_a\over
1-{\lambda p'_0}}+{q'_a\over 1-{\lambda q'_0}}
\end{equation}
This leads to a number of kinematic novelties at high energies.
For instance (if $\lambda>0$) a particle close to Planck energy
becomes more and more unreceptive to receiving energy in
collisions, no matter how hard one might hit it. This may be thought
of as a novel kind of asymptotic freedom.

\subsection{Modified Fock space}

We finally note that even though the non-linearities of our formalism
distinguish between elementary particles and composites,
its set up is purely classical.
Elementary particles, however, are quantum particles, i.e.
excitations of quantum fields for which a Fock space can be
defined. It is therefore reassuring that one can easily adapt our
construction to quantum particles.
Let the Fock space be defined in the usual way, i.e. by means of
standard creation and annihilation operators $a^\dagger(p)$ and
$a(p)$, so that the vacuum satisfies $a(p)|0\rangle=0$ for all
$p$, single particle states take the form
$|p\rangle=a^\dagger(p)|0\rangle$ and multiple particle states are
defined by iteration. We can therefore write the quantum {\it
free} hamiltonian as:
\begin{equation}
{\hat H}_0=\sum_k \hbar E_k|k\rangle \langle k| +\sum_{k,k'} (E_k
\oplus  E_{k'}) |k k'\rangle\langle k k'| + ...
\end{equation}
Quantum interactions can be written likewise. For instance a
$\phi^4$ interaction leads to the interacting hamiltonian:
\begin{equation}
{\hat H}_{int}=:\phi^4: \approx \sum_{kk'}\sum_{pp'}\delta(p\oplus
p'-k\oplus k')a^\dagger(p)a^\dagger(p')a(k)a(k')
\end{equation}
Thus we have incorporated our proposal for classical momentum
addition into a quantum framework.

\section{Modifications of the threshold anomalies}
\label{encons}

Now that we know that our theory is consistent with energy-momentum
conservation and is not obviously in contradiction with the observed
properties of macroscopic bodies, we may attempt to apply it to the
real world.  The first application we would like to consider is
to follow the suggestion of Amelino-Camelia and Piran\cite{piran} that a
mofified dispersion relation may resolve the problem of the
observed threshold anomalies.
 We study first the gamma ray anomaly, because it
is a bit simpler, then the cosmic ray anomaly.  We will see that
while it was essential to establish that energy and momentum
are conserved in the theory, the analysis is actually simpler
than might have been expected.

\subsection{Gamma ray threshold anomalies}

The issue of the gamma ray threshold anomaly arises because
one expects a cut off at around 10 Tev in
the flux of gamma rays, due to their interaction with the
infra-red background. At these energies it becomes kinematically
possible to produce an electron positron pair by scattering of
a gamma ray from a photon of the infra-red background, leading
to a prediction for an upper limit to observed energies.
However, while the experimental situation is still somewhat
controversial, there are indications that the predicted threshold
is not observed (see, eg.\cite{gammaexp}).

For a threshold reaction, in the center of mass frame the electron
and positron have no momentum. Hence, due to momentum
conservation, in this frame the gamma ray and the infra-red photon
have the same energy. Energy conservation then implies that their
energy equals the rest energy of an electron $m_e$. We can draw
these conclusions because $m_e\ll E_P$, and so all corrections
imposed by our theory are negligible.

We then need to perform a boost transformation from the center of
mass frame to the cosmological frame. This can be pinned down by
the condition that one of the photons be redshifted to the
infra-red background energy. Since in this process all energies
involved are again sub-Planckian we can use plain special
relativistic formulae to conclude that $E_{IR}=(1-v)\gamma m_e$,
and since $\gamma\gg 1$ (implying $1-v\approx 1/(2\gamma^2)$) we
have:
\begin{equation}
\gamma={m_e\over 2E_{IR}}
\end{equation}
The same boost transformation blueshifts the other photon to
our predicted value for the gamma ray threshold energy. This
operation, however, may have to be performed with the corrected
boost. The uncorrected threshold energy is
\begin{equation}
E_{th0}=\gamma(1+v)m_e\approx 2\gamma m_e={m_e^2\over E_{IR}}
\end{equation}
This is now corrected to:
\begin{equation}
E_{th}=U^{-1}(E_{th0})
\label{problem2}
\end{equation}
since the full boost is now $U(E_{th})=\gamma(1+v)U(m_e)$ and
$U(m_e)\approx m_e$.

We may now obtain exact threshold formulae for the various
proposals in the literature. For \cite{leejoao} we have
\begin{equation}\label{th1}
E_{th}={{m_e^2\over E_{IR}}\over 1+{\lambda m_e^2\over E_{IR}}}
\end{equation}
and for the $\kappa$-Poincar\'e group:
\begin{equation}\label{th2}
E_{th}=\lambda^{-1}\sinh^{-1}{\lambda m_e^2\over E_{IR}}
\end{equation}
In both cases we note that with $\lambda>0$ the threshold is {\it
lowered} rather than raised. Hence if the observations have
anything to do with these dispersion relations the implication seems to be
that $\lambda<0$. In this case the invariant Planck energy
is negative $E_P=-\lambda^{-1}$,
a situation already discussed in \cite{leejoao}.

The dispersion relation (\ref{vsl2}) on the other hand is obtained
from a $U$ acting on energy like the $U$ but with
$\lambda\rightarrow -\lambda$. It is therefore not surprising that
the threshold formula in this theory is
\begin{equation}
E_{th}={{m_e^2\over E_{IR}}\over 1-{\lambda m_e^2\over E_{IR}}}
\end{equation}
which is raised with $\lambda>0$. However, the invariant in this
case is also a negative Planck energy $E_P=-\lambda^{-1}$, so the
previous conclusion remains - threshold anomalies imply a negative
Planck energy for dispersion relations proposed in literature.
This example is interesting as it tells us that the modifications
necessary to raise the speed of light, if the theory is to serve
as a VSL theory and explain the horizon problem, are of the same
sign as those required to explain the absence of the thresholds,
at least in these classes of models.

Regardless of the issue of the sign of $E_P$ there remains its
order of magnitude. From (\ref{th1}) we get:
\begin{equation}
\lambda={1\over E_{th}}-{1\over E_{th0}}
\label{problem1}
\end{equation}
so we find that we would need $|E_P|\sim 10$ Tev to explain the
gamma ray anomaly. In addition fine tuning is required: how close
$|\lambda^{-1}|$ lies to $E_{th0}$ determines the actual threshold
energy $E_{th}$.

This example teaches us an important lesson.
So long as the modified transformation law has a single
dimensional parameter, $\mu \approx \lambda^{{-1}}$ then
from (\ref{problem1}) we see
that if the usual and new threshold are the same rough order of
magnitude, then $\mu$ must be of the same order of magnitude as well.
This problem is a direct consequence of (\ref{problem2}), from which
we can see that so long as there are no small dimensionless parameters
in $U$ then the result is a formula with three dimensional parameters;
so long as two are of the same order of magnitude, so must be the third.
This tells us that the mechanism for moving the threshold used
by Amelino-Camelia and Piran in \cite{piran} cannot work in a relativistic
theory, because it relies on a coincidence of small ratios in the
cosmological frame. However, this coincidence does
not exist in all frames of reference, hence it cannot be part of the
solution of the problem in a relativistic theory.  This is then not a
problem with our example, but a general issue with theories of the kind
we are considering, which preserve the relativity of inertial frames.

\subsection{UHECR threshold anomalies}

A similar anomaly also seems to plague ultra high energy cosmic
rays (UHECRs). These are rare showers derived from a primary
cosmic ray, probably a proton, with energy above $10^{11}$ Gev. At
these energies there are no known cosmic ray sources within our
own galaxy, so it's expected that in their travels, the
extra-galactic UHECRs interact significantly with the cosmic
microwave background (CMB). These interactions should impose a
hard cut-off above $E_{th0}\approx 10^{11}$ Gev, the energy at
which it becomes kinematically possible to produce a
pion. This is the so-called GZK cut-off; however UHECRs have been
observed beyond the threshold \cite{review,crexp}.

An argument identical to the one just made for gamma rays leads to the
conclusion that any corrections imposed in our theory appear at
the level of boosting the proton in a threshold reaction from the
center of mass to the cosmological frame. Hence the corrected
threshold formula is simply $E_{th}=U^{-1}(E_{th0})$. However a
novelty appears at this stage because the proton is not an
elementary particle, so that in the boost transformation we should
replace $\lambda$ by $\lambda / N_p$, where $N_p$ is the ``number
of quanta living inside a proton''. Hence the correct formula is
\begin{equation}
E_{th}=U^{-1}[E_{th0};\lambda / N_p](E_{th0})
\end{equation}
All formulae presented above for gamma rays may now be adapted to
UHECRs. The conclusion is now that $N_pE_P\sim -10^{11}Gev$ for
previously proposed dispersion relations. Using the more general
definition (\ref{add3}) we have $f(E_P^{-1},N_p)\sim -10^{11}Gev$.

The question is then what is the right value of $N_p$ in the case
of a confined state such as the proton. An answer to this problem
may require the application of this theory to the full quantum
field theory of $QCD$.  From the point of view of phenomenology,
the suggestion in any case is that the parameters that modify the
boost properties of the proton may differ from those of the
electron and photon. One may then adjust the free function
$f(\lambda,n)$ used in defining the multi-particle sector to
reconcile the different in energy scales of the cosmic ray and
gamma ray thresholds.  However, given that $N_p$ for the proton
is likely of order $3$, it is difficult to see how this could
be accomplished for any simple function $f$, unless it contains
small dimensionless parameters.

\subsection{How to resolve all five issues}

From the preceeding discussion we see that there is a basic problem
with using a modified form of special relativity such as we are
considering here to solve the problem of the threshold anomolies.
The problem is that so long as the function $f_{1}$ has a single
dimensional parameter, $\lambda$ and no small dimensionless parameters,
then
$\lambda^{{-1}}$ must not be too many orders of magnitude away from the
threshold predicted by the usual linear theory.  This prevents
a single kinematical effect from solving both the gamma ray and
UHECR anomalies, as they occur at very different energies, it also
prevents a theory with $\lambda$ on the order of the Planck
length from solving either.

One might conclude from this that in the event that observations
do in the end support the hypothesis that modified dispersion relations
with the approximate form of
(\ref{moddisp}), when applied in the cosmological rest frame, do
resolve the threshold anomolies, this would be inconsistent with the
relativity of inertial frames. Indeed, our argument shows that no
simply form of $f_{1}$ depending on only one scale, could resolve
the problem in a relativistic theory, so long as that scale
were the Planck scale.

However, before reaching such a drastic conclusion, there is a simpler
possibility to consider, which is that the function $f_{1}$ has
more than one scale in it.   To see that this is sufficient to
resolve the UHECR anomolie, while preserving an invariant energy
scale of the order of the Planck energy,
note that we can multiply the previous $f_1$ used in fitting the
threshold anomalies by a function that is approximately 1 for
$E\ll E_P=10^{19}$ Gev, but which diverges at $E_P$. For instance
we may take the function:
\begin{equation}
f_1={1\over (1+\lambda_1 E)(1-\lambda E )}
\end{equation}
with $\lambda_1\gg \lambda$. It is easy to see that if
$N_p\lambda_1^{-1}\approx 10^{11}Gev$ then the UHECR threshold is raised.
This theory makes the prediction that the actual UHCR threshold
should lie somewhere between its special relativity value
$E_{th0}$ and $N_pE_P$ (as $\lambda_1^{-1}$ can never be as high
as $E_P=\lambda^{-1}$).

In addition such a
theory displays an invariant positive energy, $E_P=\lambda^{-1}$
which may be of order $10^{19}$ Gev. Also the image of $U$
associated with this $f_1$ is $[0,\infty]$, so that in this theory
the threshold anomalies are consistent with the group property of
the action, and the principle of relativity.

Finally, can we pick a theory that satisfies the other criteria
we set out in the introduction? To see that this is straightforward,
note that  the
function $f_2$ does not enter the discussion of threshold
anomalies, and so the issues of VSL and of the existence of a
maximal momentum are decoupled from threshold anomalies. Instead
as we see from equation (\ref{speedc}) and the discussion at
the end of (\ref{cond}), both of these properties
are governed by $f_{3}=f_{2}/f_{1}$.

To avoid an energy dependent speed of light that so far would have
been detected in observations of gamma ray bursts, $f_{3}$ should
differ from unity only on the Planck scale.  For example, consider
$f_{3}= e^{E/E_{P}}$.  It is easy to see that this gives
a maximum momentum, equal to $E_{P}$ and a diverging speed
of light.

\section{Real space formulation}\label{real}

If Lorentz transformations are non-linear they take a different
aspect in real and momentum space. The choice of momentum space in
\cite{leejoao} is tied to the use of the Fock-Lorentz
representation, which has a large time-like invariant suitable for
identification with the Planck energy $E_P$ but not the Planck
time $t_P$. Once in momentum space one may ask how to recover a
real space formulation.

One prescription is to define space coordinates as the generators
of shifts in momentum space (this seems to be at odds with the
proposal in \cite{jur}). Because the theory is non-linear, shifts
are not pure additive constants, and may be read off from standard
shifts subject to a $U$ transformation. For \cite{leejoao}, small
energy shifts take the form:
\begin{eqnarray}
\delta E &=& (1-\lambda E)^2 \epsilon\\
\delta p_i &=& -\lambda(1-\lambda E)p_i \epsilon
\end{eqnarray}
whereas momentum shifts are:
\begin{eqnarray}
\delta E&=&0\\
\delta p_i&=&(1-\lambda E)\epsilon
\end{eqnarray}
Hence the corresponding spatial coordinates are:
\begin{eqnarray}
t&=&(1-\lambda p_0){\left((1-\lambda p_0){\partial\over
\partial p_0}
-\lambda D \right)}\\
x^i&=&(1-\lambda p_0){\partial\over \partial p_i}
\end{eqnarray}
This bears some resemblance to Snyder's non-commutative geometry
\cite{snyder}, which has
\begin {equation}
x^\mu={\partial\over \partial p_\mu}-\lambda p^\mu D
\end{equation}
(see Eqn.~9 in \cite{snyder}).  However there is an important
difference. As may be easily checked, the space and time
coordinates all commute with each other. \f [x^a, x^b ] =0 \ff

The price to pay for this is that there are now novelties in the
commutators of the space-time coordinates with energy and momenta.
Indeed:
\begin{eqnarray}
[x^i, p_j]&=&\delta^i_j(1-\lambda p_0)\label{xpcom}\\
{\left[x^0, p_i\right]}&=&-\lambda (1-\lambda p_0)p_i\\
{\left[x^0, p_0\right]}&=&(1-\lambda p_0)^2
\end{eqnarray}
This suggests that we have now an energy dependent Planck's
``constant'' since (\ref{xpcom}) implies $\hbar= 1-\lambda p_0$.
As a result for Planck energies there is no uncertainty principle
- the Planck energy is not only an invariant but it is also
apparently perfectly classical. We are currently investigating
further the implications of this proposal.

\section{Conclusions}

In this paper we have presented a general
method for implementing non-linear actions of the Lorentz group
based upon knowledge of the dispersion relations. Our results
complement those of other authors, who have studied the possibility
that the action of the lorentz transformations is modified at high
energies.  The approach taken in this paper generalizes that in
\cite{leejoao} by considering different maps $U$, which we
identified with functions $f_1$ and $f_2$ in Eqn.~(\ref{desp}). We
found that the group property is preserved if $U$ is invertible
and its image contains $[0,\infty]$. If $Ef_1$ diverges at some
finite energy $E_P$, this takes the place of an invariant Planck
energy. Careful design of $U$ may also explain the threshold
anomalies. The function $f_2$ may then be used to implement a
maximal momentum and a diverging speed of light.
Using the freedom to choose $f_{1}$ and $f_{2}$ we found
that all five requirements we listed in the introduction may
be achieved in one theory.  One, among many, examples that do
so is the following,
\f
f_1={1\over (1+\lambda_1 E)(1-\lambda E )} , \ \ \ \ \ \
f_{2}=e^{E/E_{Planck}}.
\ff
with $\lambda_{1} > > E_{Planck}^{-1}$.

We also discussed the extension of the transformation laws
to real space. We found that there is no general need for
the coordinates of space to become non-commutative. Instead,
by defining the coordinates of space to be generators of translations
in momentum space, we arrived at a commutative spacetime geometry.
While it remains for experiment to decide, we note that this approach
is closer to the spirit of general relativity, in which the local
properties of spacetime arise from the tangent space of a manifold.
It then may be close to that expected from the classical limit of
quantum gravity, according to which the Poincare invariance of
Minkowski spacetime has no fundamental significance, but is only
an accidental symmetry of the ground state of the classical limit.
Furthermore, by taking this point of view we discovered a novel
feature of the theory, which is that the effective Planck's constant
appears
to become energy dependent.

Of course there are many things still to do to investigate whether
theories of the kind discussed here have a chance to be true.  It is
important to understand whether the modifications of the energy
momentum relations predicted by loop quantum gravity in \cite{loopcalc}
are necessary consequences of that theory and, if so, whether that
theory predicts the existence of a prefered frame or a modification of
lorentz transformations preserving the relativity of inertial frames.
Of equal interest is the question of whether critical string
theory\footnote{There are also related results conceerning
non-critical string theory\cite{amelinovsl}.} can be made
consistent with deformed dispersion relations and modifications of
the action of lorentz transformations, or whether observations of
such effects would disprove critical string
theory\cite{stringlj}.  Indeed, the general question of how to
incorporate the kinds of modifications of kinematics contemplated here
and in related papers into a fully interacting quantum field theory
remains open, as does the question of how these modifications may be
incorporated into classical general relativity \cite{ahl,granik,contr,spin}.

Of course, the main motivation for studying this class of theories is
the hope that in the not too distant future astrophysical and
cosmological observations of the kind considered here will
teach us whether and how lorentz invariance is realized
at the scales relevent for quantum gravity.

\section*{ACKNOWLEDGEMENTS}

We are grateful for conversations with Stephon Alexander, Giovanni
Amelino-Camelia, Jurek Kowalski-Gilkman, Seth Major, Fotini
Markopoulou, John Moffat
and Carlo Rovelli,
which have helped us to understand better the idea proposed here.
Jo\~ao Magueijo thanks the Perimeter Institute for hospitality.  Lee
Smolin would like to thank the Jesse Phillips Foundation for support which
made his contribution to this work possible.

\end{document}